\documentclass[english]{article}
\usepackage[T1]{fontenc}
\usepackage[latin9]{inputenc}
\usepackage{geometry}
\usepackage{color}
\usepackage{float}
\usepackage{amssymb}
\usepackage{graphicx}
\usepackage{esint}
\usepackage{epstopdf}
\usepackage{epsfig}

\makeatletter

\usepackage{url}

\makeatother

\usepackage{babel}
\begin{document}

\title{Classical light vs. nonclassical light: Characterizations and interesting
applications}

\author{Anirban Pathak$^{a,}$\footnote{email: anirban.pathak@gmail.com},
Ajoy Ghatak$^{b,}$\footnote{email: ajoykghatak@gmail.com}}

\maketitle
\begin{center}
$^{a}$Jaypee Institute of Information Technology, A-10, Sector-62,
Noida, UP-201307, India \\
$^{b}$M.N. Saha Fellow, The National Academy of Sciences, India
(NASI), D42, Hauz Khas, New Delhi, India
\end{center}

\begin{abstract}
We briefly review the ideas that have shaped modern optics and have
led to various applications of light ranging from spectroscopy to
astrophysics, and street lights to quantum communication. The review
is primarily focused on the modern applications of classical light
and nonclassical light. Specific attention has been given to the applications
of squeezed, antibunched, and entangled states of radiation field.
Applications of Fock states (especially single photon states) in the
field of quantum communication are also discussed. 
\end{abstract}

\section{Introduction\label{sec:Introduction } }

Once Poincare said \emph{\textquotedblleft The scientist does not
study nature because it is useful to do so. He studies it because
he takes pleasure in it; and he takes pleasure in it because it is
beautiful. If nature were not beautiful, it would not be worth knowing,
and life would not be worth living\dots{} I mean the intimate beauty
which comes from the harmonious order of its parts and which a pure
intelligence can grasp\textquotedblright{}} (as quoted in Chapter
4 of \cite{chandrasekhar1987truth}). It is the search of this harmony
in nature that brought different theories of light. If we look back
at Newton's corpuscular theory of light, it would not be difficult
to guess that the harmony of nature revealed through his 3 laws of
motion and the law of gravity (all four of which are obeyed by particles),
which together can explain almost every phenomena known at that time
might have forced him to consider light also as corpuscular. Later
on, this search for harmony led Maxwell to discover the intimate relation
between the earlier known laws of electricity, magnetism and light
(optics). It may be noted that Maxwell's main contribution in the
famous Maxwell's equation was to modify Ampere's law by introducing
the idea of displacement current and thus to introduce a symmetry
among the laws involving electric field and magnetic field \cite{maxwell1865dynamical}\footnote{Interested readers may freely read Maxwell's original paper at http://www.jstor.org/stable/pdf/108892.pdf}.
In fact, around 1860, Maxwell summed up all the laws of electricity
and magnetism in the form of 4 equations -{}- which are now known
as Maxwell\textquoteright s equations. He showed that, in free space,
electric field $\overrightarrow{E}(z,t)=\hat{x}E_{0}\cos\left(kz-\omega t\right)$
satisfies the 4 equations with the corresponding magnetic field given
by $\overrightarrow{H}(z,t)=\hat{y}H_{0}\cos\left(kz-\omega t\right);$
$H_{0}=\sqrt{\frac{\mu_{0}}{\epsilon_{0}}}E_{0},$ where $\hat{x}$
and $\hat{y}$ represent unit vector in $X$ and $Y$ directions,
respectively and $\epsilon_{0}=8.854\times10^{-12}\,{\rm C^{2}N^{-1}m^{-2}}$
is the dielectric permittivity and $\mu_{0}=4\pi\times10^{-7}$${\rm Ns^{2}C^{-2}}$
is the magnetic permeability of the free space (classical vacuum).
The above equations describe propagating electromagnetic waves. Thus
from the laws of electricity and magnetism, Maxwell predicted the
existence of electromagnetic waves, and by substituting the above
solutions in Maxwell\textquoteright s equations he showed that the
velocity (in free space) would be given by $c=\frac{\omega}{k}=\frac{1}{\sqrt{\epsilon_{0}\mu_{0}}}\thickapprox3\times10^{8}$
m/s. Thus, Maxwell not only predicted the existence of electromagnetic
waves, he also predicted that the speed of the electromagnetic waves
in air should be about $3\times10^{8}$ m/s. He found that this value
to be very nearly equal to the measured value of velocity of light
(in air) known in that time. In fact, in 1849, Fizeau measured the
speed of light (in air) as $3.14858\times10^{8}$ m/s. The sole fact
that the two values were very close to each other led Maxwell to propound
(around 1865) his famous electromagnetic theory of light. Here, we
may note that observing a great symmetry (the fact that velocity of
electromagnetic wave and that of light are nearly the same) present
in nature, Maxwell conjectured that light is an electromagnetic wave.
In making this powerful conjecture without any available experimental
evidence, Maxwell actually showed his confidence on the fact that
nature is beautiful and symmetric. 

The confidence on the beauty of nature shown by Maxwell in particular
and scientists in general is nicely reflected in a conversation between
Einstein and Heisenberg, which was recorded by Heisenberg as \cite{chandrasekhar1987truth}-
\emph{\textquotedblleft If nature leads us to mathematical forms of
great simplicity and beauty\textemdash by forms, I am referring to
coherent systems of hypotheses, axioms, etc. (etc.,)\textemdash to
forms that no one has previously encountered, we cannot help thinking
that they are \textquotedblleft true,\textquotedblright{} that they
reveal a genuine feature of nature\dots . You must have felt this
too: the almost frightening simplicity and wholeness of the relationships
which nature suddenly spreads out before us and for which none of
us was in the least prepared.\textquotedblright{}} The simplicity
and beauty referred here were vibrantly present in Maxwell's equations
and those compelled Maxwell to consider light as an electromagnetic
wave. It was confidence on the beauty of the mathematical forms of
Maxwell's beautiful equations, which forced Einstein to show confidence
on these equations rather that on the century old and well tested
Galilean transformations\footnote{Everyday, we see that relative velocity of two cars that approach
each other with the same speed is double of the individual speed.
This is in accordance with the Galilean transformation, but according
to Maxwell's equation, light would always move with a constant velocity
$c$ in free space. Thus, if we send light from two torches in the
opposite direction, their relative velocity would still remain $c.$
This was in sharp contrast with the Galilean transformation. } and indirectly this confidence on the symmetry of the Maxwell's equations
led him to introduce the special theory of relativity. 

Historically, light played an extremely important role in understanding
nature. For example, most of the information that we have about the
celestial bodies are received through light (may not be restricted
only to the visible range). However, at a more fundamental level,
an effort to understand the blackbody spectrum (i.e., to explain experimental
observations related to intensity of lights of different wavelength
emitted by a blackbody) led Planck to postulate that energy from an
electric oscillator (which constitutes the wall of a cavity) had to
be transferred to electromagnetic waves in different quanta of each
\cite{planck1901law}, but the waves themselves would follow the conventional
wave theory of Maxwell. This was postulated in 1900\footnote{Planck's paper cited here as \cite{planck1901law} was published in
1901, but the paper contains following note- In other form reported
in the German Physical Society (Deutsche Physikalische Gesellschaft)
in the meetings of October 19 and December 14, 1900, published in
Verh. Dtsch. Phys. Ges. Berlin, (1900) \textbf{2}, 202 and 237. 

An English translation of Verh. Dtsch. Phys. Ges. Berlin, (1900) \textbf{2},
237 is available at http://hermes.ffn.ub.es/luisnavarro/nuevo\_maletin/Planck\%20(1900),\%20Distribution\%20Law.pdf}. Just after 5 years, in 1905, Einstein (while he was working at the
Swiss Patent office) published a set of five outstanding papers which,
according to John Satchel \emph{\textquotedblleft changed the face
of physics\textquotedblright{}} \cite{stachel1999einstein}. In one
of those 5 papers \cite{einstein1905erzeugung}, he introduced his
famous theory of light quanta according to which light is considered
to be consisted of mutually independent quanta of energy 
\begin{equation}
E=h\nu,\label{eq:hnu}
\end{equation}
where $\nu$ is the frequency and $h$ is the Planck\textquoteright s
constant. Here it is important to note that there was a fundamental
difference between Planck's idea of light quanta and that of Einstein.
Specifically, Planck postulated that energy from an electric oscillator
had to be transferred to electromagnetic waves in different quanta
of each, but the waves themselves would follow the wave theory of
Maxwell. In contrast, Einstein assumed that energy is not only given
to an electromagnetic wave in separate quanta, but is also carried
in separate quanta. Einstein's revolutionary idea of light quanta
explained  an interesting observation related to light. To be precise,
in 1887, Hertz did a simple experiment with light \cite{hertz1887ueber}.
In his experiment, electrodes illuminated with the ultraviolet (UV)
light were found to emit electrons. This phenomenon is known as the
photoelectric effect, and Einstein postulated ``light quantum'',
to explain this phenomenon. Thus, the revolutionary ideas of both
Planck and Einstein were theoretical in nature, but were obtained
from the efforts to explain experimental observations related to light
and these ideas subsequently played important role in the construction
of quantum physics, the best known model of nature.

Before we proceed further, it would be interesting to note that in
\cite{einstein1905erzeugung}, Einstein obtained Eq. (\ref{eq:hnu})
by comparing entropy of radiation with that of a gas having $n$ molecules.
Specifically, he had shown that if volume changes from $V_{0}$ to
$V$, then the change in entropy of radiation having a fixed amount
of total energy is given by 
\begin{equation}
S-S_{0}=k\ln\left(\frac{V}{V_{0}}\right)^{\frac{E}{h\nu}},\label{eq:einstein1}
\end{equation}
whereas the corresponding change in entropy for an ideal gas having
$n$ particles is
\begin{equation}
S-S_{0}=k\ln\left(\frac{V}{V_{0}}\right)^{n}.\label{eq:einstein2}
\end{equation}
Comparing Eqs. (\ref{eq:einstein1}) and (\ref{eq:einstein2}), Einstein
reached to the conclusion that radiation behaves in manner\textcolor{red}{{}
}like it is composed of independent light quanta and $\frac{E}{h\nu}$
should represent the total number of light quanta ($n)$ having individual
energy of $h\nu$ \cite{ghatak2017optics}. It is of further interest
to note that later on, a few scientists have tried to explain photoelectric
effect without using the concept of photon or light quanta. They assumed
that the energy of the atoms constituting the electrode on which light
falls is quantized \cite{bosanac1998semiclassical}. Thus, photoelectric
effect can be explained by considering quantization of either light
or matter. However, it seems obvious that Einstein used the concept
of light quanta. This is so because Einstein provided an explanation
of the photoelectric effect in 1905, when neither Rutherford\textquoteright s
model (1909), nor Bohr model (1913) was known\footnote{It may be noted that Bohr model also originated in an effort to explain
the origin of lights of certain wavelengths (as was observed in Lyman,
Balmer, Paschen, Bracket and Pfund series.}, but Planck\textquoteright s idea was already present since 1900.
Naturally, Einstein used the concept of light quanta in his explanation
of the photoelectric effect. This discussion establishes two points: 
\begin{enumerate}
\item It is important to know the history of a subject to understand that
subject.
\item Light played a fundamental role in the development of the most fascinating
and useful concepts of the modern physics. 
\end{enumerate}
In what follows, we would keep this in mind and would try to provide
a historical (but not chronological) overview of the development of
various concepts related to modern optics and modern applications
of them.

Maxwell's work provided a clear understanding of electromagnetic wave
which still plays the most crucial role in communication engineering
and enables us to speak with friends and relatives through cell phones,
to see different channels in TV, to do online shopping, etc. On the
other hand, the concept of photon plays a crucial role in many of
the recently proposed path-breaking applications of quantum information
processing and quantum communication, such as unconditionally secure
quantum cryptography \cite{bennett1984quantum,bennett1992quantum},
quantum teleportation \cite{bennett1993teleporting}, and dense coding
\cite{bennett1992communication,mattle1996dense}. Before, we proceed
to describe some of these applications and briefly introduce the notion
of nonclassical light, we must mention that neither Planck nor Einstein
used the term ``photon''. It was only in 1926 that the American
chemist Gilbert Lewis coined the word ``photon''. In \cite{lewis1926conservation},
Lewis wrote ``\emph{\dots it spends only a minute fraction of its
existence as a carrier of radiant energy, while the rest of the time
it remains as an important structural element within the atom. \dots I
therefore take the liberty of proposing for this hypothetical new
atom, which is not light, but plays an essential part in every process
of radiation, the name photon}''. One can easily recognize that the
term photon is used today with a different meaning. Further, we would
like to note that in 1905, Maxwell\textquoteright s electromagnetic
theory was well established and consequently, Einstein\textquoteright s
idea of the light quantum was not readily accepted (for a discussion
see \cite{einstein1979autobiographical}). In fact, even today there
are some open questions related to the wave function of photon\footnote{The main problem in defining a wave function of photon in position
space arises because of the fact that it cannot be localized in position
space as it has a definite momentum.} (\cite{bialynicki1994wave,inagaki1998physical,sipe1995photon} and
references therein) and its momentum in a medium \footnote{Interested readers may read about Abraham\textendash Minkowski dilemma
in detail to know the origin of this interesting problem. About a
century ago, Abraham and Minkowski gave two different expressions
for momentum of light in a medium. To understand the dilemma, at the
single photon level, we may note that for free space momentum of a
photon is $\hbar k,$ and it's unambigious, but for a medium having
refractive index $n$, there are two competing expressions for photon
momentum: $n\hbar k$ and $\frac{\hbar k}{n}$. Both are used, and
thus the open question is: Which one of these two expressions is correct?
Apparently, the problem arises because even in the classical optics,
there is no universally accepted definition for the electromagnetic
momentum in a dispersive medium.} (see \cite{barnett2010enigma,barnett2010resolution} and references
therein); and there are people who are not confident on the existence
of photon (interested readers may read Lamb Jr.'s article entitled,
Anti-photon \cite{lamb1995anti}, where the author claimed that ``\emph{...there
is no such thing as a photon}''). Our view is different, and we believe
that the wide domain of optics can be classified into three sub-domains-
classical optics, semi-classical optics and quantum optics \cite{fox2006quantum}.
Specifically, science of describing those phenomena which can be explained
with the help of the classical theory of light (i.e., considering
light as an electromagnetic wave) and classical theory of matter (which
does not require quantization of atomic/molecular energy levels).
Reflection, refraction, dispersion, etc., are examples of phenomena
that fall under classical optics. Whereas explanation of another set
of phenomena, like Compton effect and photoelectric effect, requires
the quantum theory of matter, but does not essentially require quantum
theory of light\footnote{It is interesting to note that Nobel laureate C V Raman, provided
a semiclassical explanation of the Compton effect in Ref. \cite{raman1928classical}.}. Such phenomena fall under semiclassical optics. Finally, there exists
a set of phenomena (like the recoil of atom on the emission of light)
which cannot be explained without considering the quantum theory for
both atom and field. Those phenomena fall under the domain of quantum
optics, and a major part of this review is dedicated to the application
of such phenomena.

In his 1905 paper on the photoelectric effect \cite{einstein1905erzeugung},
Einstein conceptualized the notion of \textquotedblleft wave particle
duality\textquotedblright , which eventually led to the development
of quantum theory. Few years later, in 1923, de Broglie, showed confidence
on the symmetry and beauty of nature by claiming that nature manifest
itself in two forms- light and matter, if one of them has a dual character,
then the other one should also have the dual character \cite{de1923ondes,de1923quanta,de1923waves}.
Believing in the inherent harmony of nature, he conjectured that Fermat's
least optical path principle of optics and the least action principle
of mechanics are manifestations of the same law as their mathematical
forms are the same. This conjecture led to the idea of matter wave
and de Broglie wavelength, which again played a very important role
in the development of quantum mechanics. The fact that de Broglie
was convinced that there was a harmony in nature, and the duality
introduced through the work of Einstein was generally true, was captured
in many of de Brogile's own statements. For example, we may quote
(cf. p. 58 of \cite{cropper1970the}) : \emph{``I was convinced that
the wave-particle duality discovered by Einstein in his theory of
light quanta was absolutely general and extended to all of the physical
world, and it seemed certain to me, therefore, that the propagation
of a wave is associated with the motion of a particle of any sort-
photon, electron, proton or any other.}''

Recognizing the harmony of nature captured in the work of de Broglie,
he was awarded the 1929 Nobel Prize in Physics. The harmony of nature
discovered by him was nicely reflected in the presentation speech
of the Chairman of Nobel Committee for Physics (1929), who said: ``\emph{Louis
de Broglie had the boldness to maintain that not all the properties
of matter can be explained by the theory that it consists of corpuscles........Hence
there are not two worlds, one of light and waves, one of matter and
corpuscles. There is only a single universe.}'' (cf. Page 26.5 of
\cite{ghatak2017optics}).

In another direction of development, in 1917, Einstein \cite{einstein1917quantentheorie}
was able to introduce famous $A$ and $B$ coefficients, which can
describe the interaction between matter and radiation field. Specifically,
the stimulated emission which governs the operation of all laser (light
amplification by stimulated emission of radiation) systems were characterized
by $B$ coefficient, whereas spontaneous emission which leads to all
the spectral lines, can be characterized using Einstein's $A$ coefficient.
Einstein used thermodynamic argument to obtain $A$ coefficient. Ten
years later, in 1927, Dirac performed quantization of the electromagnetic
field \cite{dirac1927quantum}, which is now known as the second quantization\footnote{The word ``quantum'' means discrete. In quantum mechanics, we have
Hermitian operators for all the physical observables. These operators
satisfy eigenvalue equations, where the eigenfunctions are the wave
functions. Obtained eigenvalues corresponding to any operator is discrete
and on a particular measurement, we can obtain only one of those eigenvalues
as the value of the corresponding physical observable. Thus, in quantum
mechanics, we obtain discrete values for an observable, in other words,
in quantum mechanics allowed values of physical observables get quantized.
Historically, at the beginning of quantum mechanics (say, between
1925-1926), it was restricted to the quantization of the motion of
particles, only, and in all the early works of the founder fathers
of quantum mechanics (e.g., Schrodinger, Heisenberg, Dirac), electromagnetic
field was treated classically. Later, in 1927, Dirac quantized electromagnetic
field \cite{dirac1927quantum}, subsequently, Jordan and Wigner developed
a formalism in which particles are also represented by quantized fields.
This led to quantum field theory, which has been formulated in the
language of second quantization.}. In fact, quantization of field in general and radiation field in
particular is referred to as second quantization, and it naturally
yields Einstein's $A$ coefficient and the concept of light quanta.
In the mean time, in 1924, Bose \cite{bose1924plancks} provided a
quantitative explanation of Planck's law and paved the way for quantum
statistics by introducing a technique for counting statistics of particles
having zero rest mass \cite{agarwal2012quantum}. This work of Bose
was followed by another seminal paper of Einstein, in which counting
statistics for particles having finite mass (boson) was provided.
These works are relevant here because photons or light quanta are
bosons and they follow Bose-Einstein statistics, introduced through
the works of Bose and Einstein. Later, quantization of a system of
finite rest mass was performed by the Russian physicist Vladimir Fock
\cite{agarwal2012quantum}; the corresponding space (i.e., the appropriate
state space for the electromagnetic field) is called the Fock space
and the basis states of this space are referred to as the Fock states
or number states $|n\rangle$. To be precise, for the present review,
we are only interested in bosonic Fock space, and wish to express
states of the radiation field in Fock basis. From the discussion,
so far, we can easily recognize that if one uses second-quantization
formalism and Fock basis, he can express an arbitrary radiation field
state as $|\psi\rangle=\sum_{n=0}^{\infty}c_{n}|n\rangle,$ where
$|n\rangle$ represents a Fock state, more lucidly $|0\rangle$ corresponds
to vacuum state, $|1\rangle$ corresponds to a single photon state,
$|n\rangle$ corresponds to a state with $n$ photons, $|c_{n}|^{2}=P(n)$
is the probability of obtaining $n$ photons ($P(n)$ is also referred
to as the photon number distribution) if the number of photons present
in the quantum state $|\psi\rangle$ is measured. Clearly, in this
formalism (formalism of second quantization), notion of light quanta
follows, automatically. However, that's not our concern. Our concern
is now, as the electromagnetic field is quantized in general, and
as every state of the radiation field is essentially quantum because
it can always be described as a quantum state $|\psi\rangle=\sum_{n=0}^{\infty}c_{n}|n\rangle,$
how to distinguish classical and quantum light? Here, we need to come
out of the popular classification made by using particle nature and
wave nature of light and note that there are some properties of quantum
world which are not present in the classical world. For example, in
the quantum world, one cannot measure two non-commuting operators
(that represent two physical observables) simultaneously with arbitrary
accuracy. This is known as Heisenberg's uncertainty principle. No
such, uncertainty exists in the classical world, so a quantum state
can be approximated as classical if the observed uncertainty (associated
with both the noncommuting operators) reaches a minimum possible value
for that state. In some sense, in a world where every state is quantum,
such a quantum state can be viewed as the most classical state (or
a state which is closest to a classical state). Let's now translate
this scenario into the context of the radiation field. Traditionally,
when we look at a plane wave (a solution of Maxwell's equations),
the amplitude of the wave ($E_{0}$) is considered as a complex number,
real and imaginary parts of which are referred to as the in-phase
and out-of-phase quadratures of the field \cite{agarwal2012quantum}.
In the domain of quantum mechanics, $E_{0}$ is replaced by an annihilation
operator $a$ for that mode and the corresponding field quadratures
are defined as $X=\frac{1}{\sqrt{2}}\left(a+a^{\dagger}\right)$ and
$Y=-\frac{i}{\sqrt{2}}\left(a-a^{\dagger}\right).$ Clearly, $X$
and $Y$ don't commute as $[a,a^{\dagger}]=1.$ In fact, using $[a,a^{\dagger}]=1,$
we obtain $[X,Y]=i$. Thus, these field quadratures which correspond
to measurable quantities (i.e., physical observables) don't commute
and consequently cannot be measured simultaneously with arbitrary
accuracy. Specifically, we obtain an uncertainty relation involving
the fluctuations in the field quadrature as 
\begin{equation}
\Delta X\Delta Y\geq\frac{1}{2}.\label{eq:uncertainty}
\end{equation}
In the above discussion, we have assumed $\hbar=1.$ Now, we know
that no such uncertainty exists for the classical field, and in principle,
one can perform homodyne measurement and simultaneously measure field
quadratures with arbitrary accuracy, but quantum mechanics does not
allow that. This led to a new question: How close a quantum state
can be to the states of the classical world, where there was no uncertainty.
The quantum state of light closest to classical world of no-uncertainty
would definitely be the one with minimum uncertainty, i.e., a state
of radiation field which would satisfy $\left(\Delta X\right)^{2}=\left(\Delta Y\right)^{2}=\frac{1}{2}.$
Such a state is called coherent state, which will be elaborated separately
in the next section. Coherent states and their statistical mixtures
are considered as classical states of radiation field (classical light),
and all other states of radiation field are referred to as nonclassical
states (nonclassical light). A more formal definition of nonclassical
states will be given in the next section, but before that we may just
note that the lucid classification of light made earlier, is consistent
with this modern view. This is so because, light quanta of Einstein
can be viewed as Fock state, and for a Fock state $|n\rangle,$ we
obtain $\left(\Delta X\right)^{2}=\left(\Delta Y\right)^{2}=n+\frac{1}{2},$
which clearly indicates that, except vacuum state $|0\rangle$, no
Fock state gives us minimum uncertainty states. Further, all Fock
states (except $|0\rangle)$, $(\Delta N)^{2}=0<\bar{N}=n$ and thus
they show sub-Poissonian photon statistics (which is a signature of
nonclassicality- cf. Sec. \ref{sec:Coherent-states-and} for a relatively
elaborate discussion) and are nonclassical states (in other words
they are quantum states having no classical analogue), whereas a vacuum
state can be considered as a classical state. Similarly, electromagnetic
fields for which field quadratures can be measured with accuracy are
definitely classical. Now, we may further stress on this point by
noting that in the framework of quantum mechanics, every state is
a quantum state. As a consequence, the so called classical states
are also quantum, and need to obey no go theorems of quantum mechanics,
like Heisenberg's uncertainty principle. However, for a classical
state, the uncertainty would be minimum. Thus, in the framework of
quantum mechanics a classical state would mean a state closest to
classical world (where there is no uncertainty), in the sense that
the uncertainty in the measured values of two noncommuting observables
(for us two quadratures of the field) would be minimum for them. However,
a state that satisfy Eq. (\ref{eq:uncertainty}), may have reduced
small fluctuations (reduced with respect to the coherent state value)
in one of the quadratures at the cost of increased fluctuations in
the other quadrature. Such a state is referred to as a squeezed state.
For example, any state of radiation field that would satisfy $(\Delta X)^{2}<\frac{1}{2}$
or $(\Delta Y)^{2}<\frac{1}{2}$ would be referred to as a squeezed
state, and all squeezed states are nonclassical. We will further elaborate
on squeezed states and their applications in Sec. \ref{subsec:Squeezed-state-and}.
Keeping this distinction between classical light and non-classical
light in mind, in what follows, we will first provide a more formal
definition of classical and nonclassical states of light and then
state various modern applications of both types of light.

The rest of the paper is organized as follows. In Sec. \ref{sec:Coherent-states-and},
we formally introduce coherent state and the notion of nonclassical
states and the Glauber-Sudarshan $P$-function. In Sec. \ref{sec:Nonlinear-optics-and},
a set of interesting nonlinear optical phenomena and their applications
are discussed. Sec. \ref{sec:Characterization-of-nonclassical} is
dedicated to the methods that are used to identify nonclassical light.
In Sec. \ref{sec:Applications-of-nonclassical}, applications of nonclassical
states of radiation field (i.e., nonclassical light) are discussed
with a specific focus on the applications of squeezed, antibunched
and entangled states of light and the recent developments in the field
of quantum state engineering and quantum information processing in
general and quantum communication in particular. In Sec. \ref{sec:Applications-of-classical},
the discussion on the modern applications of light is continued, and
the modern applications of classical light are reviewed. Finally,
the paper is concluded in Sec. \ref{sec:Conclusion} with a brief
mention of some classical and nonclassical light-based technologies
that may appear in the near future.

\section{Coherent states and the idea of classical and nonclassical states
of radiation field\label{sec:Coherent-states-and} }

Let us now formally define a coherent state. For this review, we may
consider a coherent state $|\alpha\rangle$ as a state of the radiation
field, which is defined as an eigenket of annihilation operator $a$
(thus, $a|\alpha\rangle=\alpha|\alpha\rangle$ defines a coherent
state). A coherent state can also be defined using two other equivalent
definitions. Specifically, as a displaced vacuum state or a minimum
uncertainty state (as mentioned in the previous section). In infinite
dimensional Hilbert space, these definitions are equivalent\footnote{It may be noted that for the finite dimensional Hilbert space, these
definitions are not equivalent, and any finite superposition of Fock
states is always nonclassical. } and lead to a well defined state which can be expanded in terms of
Fock basis $\left\{ |n\rangle\right\} $ (introduced in the previous
section) as $|\alpha\rangle=\sum_{n=0}^{\infty}\frac{\alpha^{n}\exp\left(-\frac{|\alpha|^{2}}{2}\right)}{\sqrt{n!}}|n\rangle=\sum_{n=0}^{\infty}c_{n}|n\rangle,$
where $|n\rangle$ represents a Fock state and $\bar{N}=\langle\alpha|a^{\dagger}a|\alpha\rangle=|\alpha|^{2}$
is the average photon number. Looking at the functional form of the
probability distribution defined by $P(n)=|c_{n}|^{2},$ one can identify
that the photon number distribution for the coherent state of light
is Poissonian\footnote{In our notation, a Poissonian distribution is one which follows $P(n)=\frac{\bar{N}^{n}}{n!}\exp\left(-\bar{N}\right)$
and $(\Delta N)^{2}=\bar{N}$. Here, $|\alpha\rangle$ can be easily
recognized as a coherent state by noting that $\bar{N}=|\alpha|^{2}.$}, and it would satisfy $(\Delta N)^{2}=\bar{N}.$ If a state satisfies
$(\Delta N)^{2}<\bar{N}$, then the state will be referred to as a
sub-Poissonian state and such a state will be nonclassical. Before
we elaborate on other nonclassical states, let us first define nonclassicality.

Now, we may note that in quantum mechanics, a pure state is either
described through its wave function $|\psi\rangle$ or through its
density matrix $\rho=|\psi\rangle\langle\psi|$. However, if two pure
states $|\psi_{1}\rangle$ and $|\psi_{2}\rangle$ are mixed with
probability $p_{1}$ and $p_{2}$ then the density matrix of the state
would be $\rho^{\prime}=p_{1}|\psi_{1}\rangle\langle\psi_{1}|+p_{2}|\psi_{2}\rangle\langle\psi_{2}|=p_{1}\rho_{1}+p_{2}\rho_{2}\,\,:p_{1}+p_{2}=1.$
Thus, in general, density matrix of a mixed state $\rho$ would be
$\rho=\sum_{i=1}^{N}p_{i}\rho_{i}:\,\,\sum_{i=1}^{N}p_{i}=1$. Now,
consider a state which is a mixture of coherent states $|\alpha\rangle$,
then we must have $\rho=\int P(\alpha)|\alpha\rangle\langle\alpha|d^{2}\alpha,$
where the summation has been replaced by integration considering $\alpha$
as a continuous variable, and discrete probability $p_{i}=p_{\alpha}$
is replaced by a probability distribution $P(\alpha):\,\,\int P(\alpha)d^{2}\alpha=1$.
Now, for a mixture of coherent states $P(\alpha)$ must be nonnegative
(i.e., $P(\alpha)\geq0\,\forall\alpha)$ and must satisfy $\int P(\alpha)d^{2}\alpha=1$.
In that case, we would say that $P(\alpha)$ is a true probability
distribution. 

Coherent states form an over complete basis as for any two coherent
state $|\alpha\rangle$ and $|\beta\rangle,$ we obtain $\langle\alpha|\beta\rangle\neq\delta(\alpha-\beta).$
Thus, we may diagonally expand any quantum state\footnote{Note that this description is valid for any quantum state and it's
not restrcited to the quantum states of radiation field.} $\rho$ in the coherent state basis as 
\begin{equation}
\rho=\int P(\alpha)|\alpha\rangle\langle\alpha|d^{2}\alpha.\label{eq:pfn}
\end{equation}
However, in this expansion, $P(\alpha)$ which is usually referred
to as Glauber-Sudarshan $P$-function\footnote{Although it is usually referred to as Glauber-Sudarshan $P$-function,
and the related formulation as the Glauber-Sudarshan $P$-representation
and Glauber won 2005 Nobel prize in Physics for developing this formalism,
it is a bit controversial. Many scientists and Sudarshan himself often
argue that this representation that provide correct quantum mechanical
theory of optical coherence was actually developed by Sudarshan, and
was later adopted by Glauber, who coined the term $P$-representation.
As $P$-representation or diagonal representation played crucial role
in the development of the non-classical optics, this debate about
the origin of $P$-representation is in existence since long. However,
it resurfaced in 2005-06, when Glauber won the Nobel prize in Physics
for this formulation, but Sudarshan missed it and wrote a strong letter
of objection to the Nobel committee (for a short description of the
controversy, interested readers may see \cite{sudarshan-debate1,sudarshan-debate2,sudarshan-debate3}).
To us it appears that Nobel committee gave more credit to Glauber's
1963 paper \cite{glauber1963photon} published in February 1963, over
Sudarshan's more powerful work \cite{sudarshan1963equivalence} published
in April 1963. However, $P$-representation or diagonal representation
(or, equivalently optical equivalence theorem) was actually developed
by Sudarshan and it would have been more appropriate it call it Sudarshan
diagonal representation or sudarshan's $\phi$-representation as he
had used $\phi(z)$ in place of $P(\alpha)$ in his pioneering work.
In fact, in Eq. (4) of Ref. \cite{sudarshan1963equivalence}, Sudarshan
expressed density function $\rho$ as $\rho=\int d^{2}z\phi(z)|z\rangle\langle z|$
where he considered $|z\rangle$ as quantum state. Almost five months
later, in Sec. VII of Ref. \cite{glauber1963coherent}, Glauber reintroduced
diagonal representation of Sudarshan as $P$-representation. Note
that Eq. (7.6) of \cite{glauber1963coherent} is the same as Eq. (\ref{eq:pfn})
given above. For a clear and chronological description of the events
that happened in 1963, see \cite{simon2009sudarshan}.} is not restricted to follow $P(\alpha)\geq0\,\forall\alpha,$ and
thus to remain a true probability distribution. To be specific a negative
value of $P$-function would mean that $P(\alpha)$ cannot be viewed
as a true probability distribution, and the corresponding state $\rho$
cannot be expressed as a mixture of coherent (classical) states. This
is why $P(\alpha)$ is often referred to as quasi-probability distribution,
and we usually say that a state which cannot be expressed as a mixture
of coherent states is nonclassical. Such nonclassical states are often
seen in the radiation field, and nonclassical states of the radiation
are the states of our interest as they don't have any classical analogue.
In what follows, any radiation field state with negative value of
$P(\alpha)$ for some $\alpha$ will be called nonclassical light,
whereas the rest will be considered as classical light. Here it would
be apt to note that the diagonal representation can be considered
as a valid representation iff an inversion formula exist \cite{simon2009sudarshan}.
Interestingly, in Eq. (6) of the pioneering work of Sudarshan \cite{sudarshan1963equivalence},
an explicit expression for $P(\alpha)$ (in Sudarshan's notation $\phi(z)$)
in terms of density matrix $\rho$ was given. Further, Sudarshan established
that the expectation value of any normal ordered operator $O=a^{\dagger k}a^{l}$
(i. e. , the operators are ordered in such a way that all creation
operators appear in the left and all the annihilation operators appear
in right), in the statistical state represented by density matrix
represented in the diagonal form given in (\ref{eq:pfn}), would be
\begin{equation}
Tr(\rho O)=Tr\left(\rho a^{\dagger k}a^{l}\right)=\int P(\alpha)\alpha^{*k}\alpha^{l}d^{2}\alpha.\label{eq:sudarshan}
\end{equation}
Great importance of this result was recognized by Sudarshan. Immediately
after introducing this result in Ref. \cite{sudarshan1963equivalence},
he wrote about Eq. (\ref{eq:sudarshan}) (notation is changed here
for the consistency), ``This is the same as the expectation value
of the complex classical function $\alpha^{*k}\alpha^{l}$ for a probability
distribution $P(\alpha)$ over the complex plane. The demonstration
above shows that any statistical state of the quantum mechanical system
may be described by a classical probability distribution over a complex
plane, provided all operators are written in the normal ordered form.
In other words, the classical complex representations can be put in
one-to-one correspondence with quantum mechanical density matrices.''
This lines describes optical equivalence theorem- probably the most
important result of quantum optics, or more precisely of nonclassical
optics. This is so because Sudarshan showed that all nonclassicalities,
if any, of a given state $\rho$ are fully captured in the departure
of the corresponding $P(\alpha)$ from being a genuine classical probability
\cite{simon2009sudarshan}. Thus, negativity of $P$-function appeared
as the defining criterion for nonclassicalty.

Negativity of $P$-function being the defining criterion for nonclassicalty,
it is both necessary and sufficient. However, $P$-function cannot
be measured experimentally\footnote{There exists an interesting paper by Kiesel et al., \cite{kiesel2008experimental}
in which experimental determination of a well-behaved $P$-function
is reported for a single-photon added thermal state. However, the
method cannot be generalized as $P$-functions of nonclassical states
are not always well-behaved. Further, to the best of our knowledge
this is the only work that reports experimental determination of $P$-function. }, and as a consequence, over time several operational criteria for
nonclassicality have been developed (for a systematic discussion on
various criteria and a long list of criteria see \cite{miranowicz2010testing}.
This list was obtained by generalizing the moment based criteria of
nonclassicality in general \cite{richter2002nonclassicality,shchukin2005nonclassical}
and entanglement in particular \cite{shchukin2005inseparability,miranowicz2009inseparability}.
A finite set of moment-based criteria for nonclassicality can only
serve as a witness of nonclassicality, while a sufficient and necessary
condition would require satisfaction of an infinite set of such nonclassicality
criteria \cite{richter2002nonclassicality,miranowicz2015statistical}.
In what follows, we will briefly mention some of these criteria.

The coherent states and the nonclassical states (such as squeezed
states) generated through the time evolution of an initial coherent
state in some physical Hamiltonian have many applications. Most of
these applications and excitement connected to them started in 1960s
after the discovery of laser and the initial excitement continued
until 1980s. However, coherent state and squeezed state were known
to the founding fathers of quantum mechanics (for an excellent review
see \cite{nieto1997discovery}, where the author describes a short
history of the discovery of coherent state and squeezed state). Just
like Einstein's miraculous year, Schrodinger also had a miraculous
year, it was 1926, first half of this year was extremely productive
for him, and he submitted 6 famous papers in this period. In one of
those papers \cite{schrodinger1926stetige}, he discovered coherent
state while he was looking for classical like states that satisfy
the minimum uncertainty condition. Just in the next year, squeezed
state was discovered by Kennard (see Sec. 4C of \cite{kennard1927quantenmechanik})\footnote{Although, coherent state and squeezed state were discovered in the
early years of quantum mechanics, their importance was realized much
later. Consequently, Schrodinger and Kennard did not receive much
credit for these discoveries. In this context, Nieto made following
very interesting remark in \cite{nieto1997discovery}- \emph{``To
be popular in physics you have to either be good or lucky. Sometimes
it is better to be lucky. But if you are going to be good, perhaps
you shouldn\textquoteright t be too good}.''}. 

Consider an arbitrary state of electromagnetic field $\rho,$ which
can be expressed in the coherent state representation as shown in
Eq. (\ref{eq:pfn}). Photon number distribution of this state would
be $P(n)=\langle n|\rho|n\rangle=\int P(\alpha)\langle n|\alpha\rangle\langle\alpha|n\rangle d^{2}\alpha=\int P(\alpha)\left|\langle n|\alpha\rangle\right|^{2}d^{2}\alpha.$
Now, since $\left|\langle n|\alpha\rangle\right|^{2}>0,$ if $P(\alpha)$
is a true probability distribution (i.e., if $P(\alpha)$ has nonnegative
values for all $\alpha$ and $\int P(\alpha)d^{2}\alpha=1)$ then
$P(n)$ must be a positive quantity. In other words $P(n)=0$ would
imply negative value of $P(\alpha)$ for some value(s) of $\alpha.$
Thus, $P(n)=0$ for some values of $n$ (which refers to a hole in
the photon number distribution) is actually a signature of nonclassicality.
The process of creating holes in the photon number distribution is
known as hole burning \cite{escher2004controlled}. Various mechanisms
for hole burning have been proposed in the recent past \cite{baseia1999note,gerry2002hole,avelar2005controlled,escher2004controlled,baseia1998hole}. 

\begin{figure}
\begin{centering}
\includegraphics[scale=0.7]{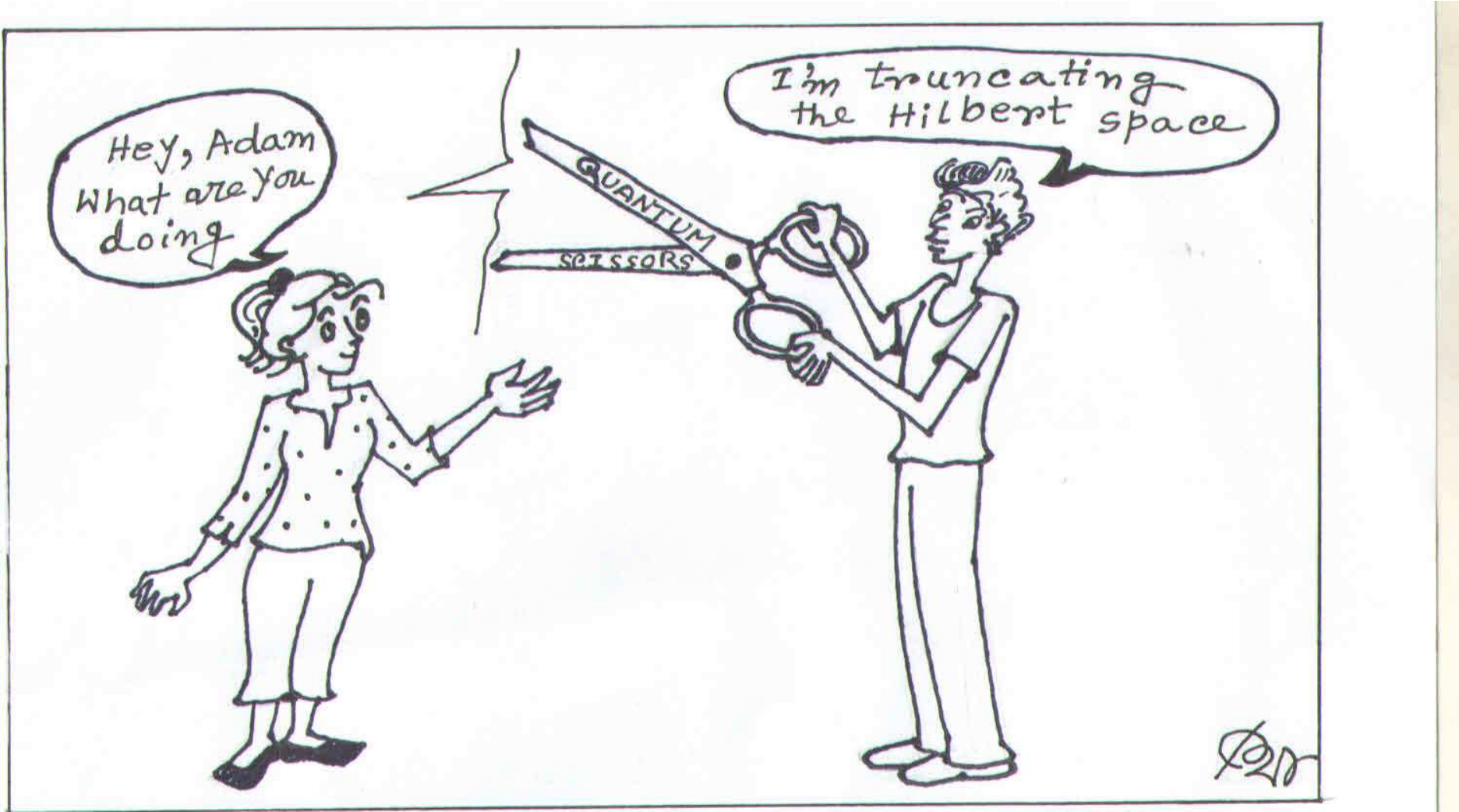}
\par\end{centering}
\caption{\label{fig:quantum-scissors}A cartoon depicting the role of quantum
scissors in the quantum state engineering, where quantum scissors
are used as devices to truncate the usual infinite dimensional Hilbert
space to $N$ dimension (where $N$ is finite) and thus to create
nonclassical state.}
\end{figure}

Let us now, look at a finite superposition of Fock states, say a quantum
state $|\psi\rangle=\sum_{n=0}^{N}c_{n}|n\rangle.$ Clearly, for this
state $P(n)=|c_{n}|^{2}$ would describe the probability of finding
an $n$ photon state. In this case, $P(n)=0\forall n>N,$ and we may
thus view it as there are a large number of holes in the photon number
distribution. This leads us to the conclusion that a finite superposition
of Fock states is always nonclassical. Procedures adopted for hole
burning and/or creation of finite dimensional states are in the heart
of quantum state engineering, which we would elaborate separately.
From the above logic it is clear that different realizations of the
finite dimensional coherent states \cite{miranowicz1994coherent,leon1997finite}
must be nonclassical. Similarly, $m$ photon added coherent state
(PACS) introduced by Agarwal and Tara \cite{agarwal1991nonclassical}
and experimentally realized (for $m=1)$ in \cite{zavatta2004quantum}
must also be nonclassical, as after the addition of one photon ($m$
photons) to every Fock states, including vacuum, we must have $P(0)=0\,\,$$\left(P(n)=0\forall n<m\right).$
It is interesting to note that the procedure of obtaining a nonclassical
state (PACS) by adding a single photon to a classical state (coherent
state $|\alpha\rangle$) manifests one of the simplest procedures
that describes classical to quantum transition. Further, PACS and
similar states are often referred to as the intermediate states \cite{dodonov2003theory,verma2008higher,verma2009reduction,verma2010generalized}
as they reduce to different well known states at different limits.
In particular, a PACS is intermediate between a fully quantum single
photon Fock state $|1\rangle$ and a coherent state $|\alpha\rangle.$
Other popular intermediate states that show nonclassical characters
at different limits are binomial state \cite{stoler1985binomial,vidiella1994statistical},
reciprocal binomial state \cite{moussa1998generation}, various types
of generalized binomial state \cite{fu1996generalized,roy1997generalized,fan1999new},
negative binomial state \cite{barnett1998negative}, excited binomial
and negative binomial states \cite{obada2002odd,wang2000excited},
hypergeometric state \cite{fu1997hypergeometric}, negative hypergeometirc
state \cite{fan1998negative}. Among these states, except negative
binomial state \cite{barnett1998negative}, all the states are finite
dimensional and naturally show nonclassicality. In addition, negative
binomial state is defined as 
\begin{equation}
|\eta,M\rangle=\sum_{n=M}^{\infty}C_{n}\left(\eta,M\right)|n\rangle,\label{eq:pacs}
\end{equation}
where $C_{n}\left(\eta,M\right)=\left[\left(\begin{array}{c}
n\\
M
\end{array}\right)\eta^{M+1}(1-\eta)^{n-M}\right]^{\frac{1}{2}},$ $0\leq\eta\leq1$ and $M$ is a nonnegative integer. Clearly, for
a nonzero $M,$ $P(0)=P(1)=\cdots P(M-1)=0,$ and these holes in photon
number distribution would imply that all negative binomial states
are nonclassical for $M\neq0.$ Thus, the fact that every finite superpostion
of Fock states is nonclassical, implies that the nonclassicalities
reported in various intermediate states \cite{verma2008higher,verma2009reduction,verma2010generalized,miranowicz2014phase,vidiella1994statistical,obada2002odd,fu1997hypergeometric,fan1998negative,pathak2014wigner}
are not surprising. Rather, they are manifestation of the above discussed
facts. Finally, a quantum scissors \cite{miranowicz2014phase,miranowicz2004dissipation}
which can be used to truncate the usual infinite dimensional Hilbert
space to a finite dimensional space must lead to nonclassicality (cf.
Fig. \ref{fig:quantum-scissors}).

\section{Nonlinear optics and applications of nonlinear optical phenomena\label{sec:Nonlinear-optics-and} }

In 1960, Theodore H. Maiman built the first laser at Hughes Research
Laboratories \cite{maiman1960stimulated}. It was a ruby laser, in
which ruby was used as the active medium to produce stimulated emission
at 694.3 nm. The realization was based on a theoretical work by Arthur
Leonard Schawlow and Charles Hard Townes \cite{schawlow1958infrared}.
The advent of ruby laser was followed by the advent of other lasing
systems, including He-Ne laser, ${\rm CO_{2}}$ laser, semi-conductor
lasers, etc. The advent of laser also contributed highly in the development
of fiber optics and experimental quantum optics. However, in this
section we will restrict ourselves to the discussion on nonlinear
optics only.

Lasing increases the intensity of light, and the output of the laser
does not diverge. Thus, the advent of lasers allowed us to apply extremely
high electric field (of the order of $10^{6}$ volts/m) to a medium
and to investigate the effect of propagation of the intense electromagnetic
wave (laser) through a medium. Such investigation led to the birth
of a new field of optics, known as nonlinear optics, where due to
the high intensity of the incident wave the linear relation between
polarization (dipole moment per unit volume) and the applied electric
field gets modified and we obtain a nonlinear relation. In fact, the
first experiment that clearly demonstrated a nonlinear optical phenomenon
was performed only after 1 year of the realization of the laser by
Maiman. Specifically, in 1961, Franken, Hill, Peters, and Weinreich
at the University of Michigan reported generation of light of wavelength
347 nm, when the output of a ruby laser (694 nm) was incident on a
quartz crystal \cite{franken1961generation}. Thus, light of frequency
$2\omega$ was generated from the incident light having frequency
$\omega.$ This process is known as second harmonic generation, and
it defines a typical nonlinear optical phenomenon as in the normal
situation (in the regime of linear optics) wavelength of the incident
light would not have changed. This process can be used to generate
blue light by passing a red laser beam through a nonlinear crystal.
This often happens inside the blue laser pointers. The presence of
a small quadratic term in the optical polarizability of a nonlinear
optical crystal led to second harmonic generation, soon after demonstrating
the second harmonic generation, the same group of scientists recognized
that this small quadratic term in the optical polarizability would
also lead to mixing of light from two different sources with two different
frequencies \cite{bass1962optical}. In other words, if we send two
light waves at two frequencies $\omega_{1}$ and $\omega_{2}$, then
the crystal can mix these two frequencies to generate light of frequency
$\omega_{1}+\omega_{2}$ (known as sum frequency generation) and/or
$\omega_{1}-\omega_{2}$ (known as difference frequency generation).
The process is now usually referred to as frequency mixing, but in
the pioneering work of Bass et al. \cite{bass1962optical}, in which
sum frequency generation was experimentally demonstrated in 1962,
it was referred to as optical mixing. It is easy to recognize that
second harmonic generation is a special case of sum frequency generation
where $\omega_{1}=\omega_{2}.$ Similarly, we can visualize third
or higher harmonic generation process as a nonlinear optical process
where higher harmonics are generated by frequency mixing. Frequency
mixing process is often used to convert frequency of a given light
to the region 800 nm-1000 nm where detectors perform with highest
efficiency. The applicability of frequency mixing in general and second
harmonic generation in particular is huge. For example, second harmonic
generation imaging microscopy has been used in the diagnostics of
diseases \cite{campagnola2011second}, imaging cells and extracellular
matrix in vivo \cite{zoumi2002imaging} and in determination of ovarian
and breast cancers \cite{tilbury2015applications}. Further, we would
like to mention another interesting nonlinear optical process- subharmonic
generation, in which a stronger beam produces two beams of frequencies
lower than the original beam. A particular case of subharmonic generation
is spontaneous parametric down conversion (SPDC) process \cite{pathak2016optical}.
Here, it would be apt to note that the nonlinear optical phenomena
may happen in both classical and quantum worlds. \footnote{Classical nonlinear optics is discussed very frequently and can be
found in many text books. To obtain an idea of quantum nonlinear optics
interested readers may look at \cite{hanamura2007quantum,peyronel2012quantum,chang2014quantum}
and references therein.}. In other words, at the output of a nonlinear optical crystal, one
may obtain classical or non-classical light depending upon other conditions
of the experiment. Specifically, type I and type II SPDC processes
are primarily used to yield entangled states of light, which have
been successfully used in realizing various ideas of quantum information
processing and quantum communication. Considering its wide applicability,
entangled states\footnote{An entangled state is a quantum state of a composite system which
cannot be expressed as a tensor product of the component systems (sub-systems)
that constitute the composite system. Specifically, if the composite
state $|\psi\rangle_{AB}\neq|\psi\rangle_{A}\otimes|\psi\rangle_{B}$,
where $|\psi\rangle_{A}$ and $|\psi\rangle_{B}$ represent arbitrary
states of subsystem $A$ and $B$, then $|\psi\rangle_{AB}$ is considered
to be entangled, otherwise it is called separable. Thus, a two photon
state $|\psi\rangle_{AB}=\frac{|HH\rangle_{AB}+|VV\rangle_{AB}}{\sqrt{2}}$
is entangled, but the state $|\psi\rangle_{AB}=\frac{|HH\rangle_{AB}+|VH\rangle_{AB}}{\sqrt{2}}=\frac{|H\rangle_{A}+|V\rangle_{A}}{\sqrt{2}}\otimes|H\rangle_{B}$
is separable. Here $|H\rangle$ and $|V\rangle$ denote horizontal
and vertical states of polarization, respectively.} are discussed separately in Sec. \ref{subsec:Entangled-state-and}.
Here, we just note that in the SPDC process of type I two nonlinear
crystals are used in such a way that the photons generated from these
two crystals are in orthogonal polarization and therefore the down
conversion occurred in either of the crystals produces entangled photons
of same polarization; whereas\textcolor{red}{{} }type II SPDC process
uses a single nonlinear crystal and entangled photons of orthogonal
polarization are generated (for an elaborate discussion see \cite{pathak2016optical}).
Before, we proceed to describe the applications of entangled states,
we would like to mention about another common nonlinear optical phenomena
known as four wave mixing (FWM). In the quantum description of the
FWM process, simultaneous annihilation of two pump photons (which
may have different frequencies) creates a signal-idler photon pair.
This nonlinear optical phenomenon is of particular interest as its
applications have been reported in various contexts (\cite{dutt2015chip,reimer2014integrated,glasser2012stimulated,wu2004ultraviolet,fiorentino2002all,ding2014observation,agha2012low,zhang2013coherent}
and references therein). Specifically, its applications are reported
for optical parametric oscillators (OPOs) \cite{dutt2015chip}, optical
filtering \cite{ding2014observation}, low noise chip-based frequency
converter \cite{agha2012low}, single-photon sources for quantum cryptography
\cite{reimer2014integrated,wu2004ultraviolet,fiorentino2002all},
frequency-comb sources \cite{reimer2014integrated}, stimulated generation
of superluminal light pulses \cite{glasser2012stimulated}, etc. Further,
several useful optical phenomena (e.g., wavelength conversion, signal
regeneration and tunable optical delay) have been observed in silicon
nanophotonic waveguides using FWM (see \cite{liu2010mid} and references
therein). FWM microscopy is also used recently to study the nonlinear
optical responses of nanostructures \cite{wang2011four}.  

With the advent of quantum information processing the challenge is
to perform nonlinear optical operations with a few photons or a low
intensity. This is so because the optical realization of ${\rm CNOT}$
and other similar quantum gates require nonlinearity, but quantum
information processing is performed with single photon or a few photons.
This is challenging because, conventional nonlinear optics is useful
only when the incident beam is sufficiently intense. The requirement
led to studies on nonlinear optics with low intensity sources \cite{harris1999nonlinear}
and a method to circumvent the problem by using linear optical elements
and a set of detectors (KLM approach) \cite{knill2001scheme}. The
KLM approach works because the quantum measurement itself is a nonlinear
process.\textcolor{red}{{} }

It is out of the scope of the present work to review all the nonlinear
optical phenomena and their applications. However, we would like to
mention that following nonlinear optical phenomena deserve special
attention because of their applications listed against their names. 
\begin{enumerate}
\item Optical parametric amplification (OPA) has applications in linear
optical amplifier, transparent wavelength conversion, return-to-zero
(RZ)-pulse generation, all-optical limiters, etc., (see \cite{hansryd2002fiber}
for a review).
\item Optical parametric oscillation (OPO) has applications in quantum noise
reduction \cite{fabre1989noise}, frequency conversion \cite{johnson1995narrow},
twin-beam generation \cite{gao1998generation}, etc. 
\item Optical rectification (OR) has applications in generation of tetrahertz
pulses \cite{schneider2006generation}, 
\item Optical Kerr effect, has applications in optical pulse compression,
mode locking of lasers, nonlinear intensity-dependent discriminator's,
picosecond time-resolved emission and absorption spectroscopy \cite{sala1975optical}.
\item Self-phase modulation (SPM) has applications in designing schemes
for all-optical data regeneration \cite{mamyshev1998all}.
\item Cross-phase modulation (XPM) has applications in quantum computation
\cite{shapiro2007continuous}, optical switching \cite{larochelle1990all},
etc. 
\item Cross-polarized wave generation (XPW) has applications in the designing
of efficient temporal cleaner for femtosecond pulses \cite{jullien2006highly}. 
\item Optical phase conjugation has applications in adaptive optics, lens-less
imaging, phase-conjugate resonators, image processing, associative
memory,\cite{giuliano1981applications,pepper1986applications}. 
\end{enumerate}

\section{Characterization of nonclassical light\label{sec:Characterization-of-nonclassical} }

\begin{figure}
\begin{centering}
\includegraphics[scale=0.6]{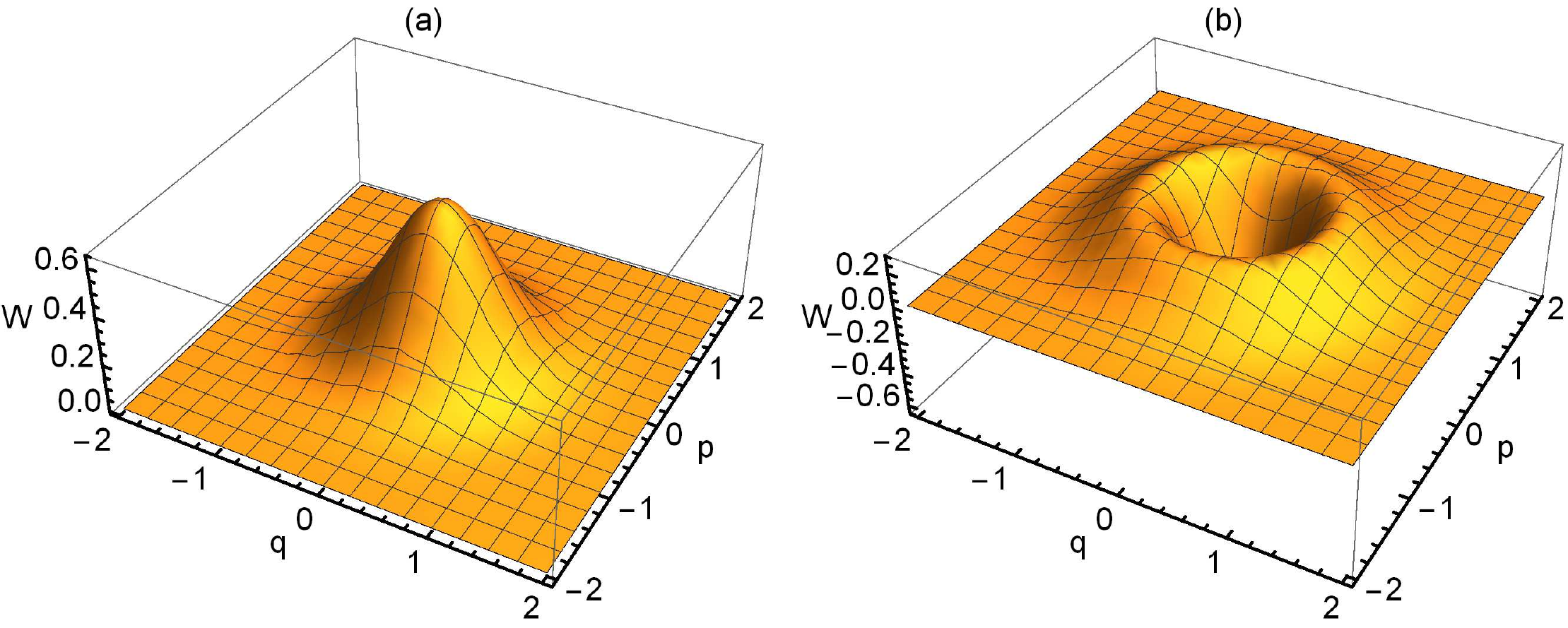}
\par\end{centering}
\caption{\label{fig:fig2wigner}(Color online) Wigner function of (a) coherent
state and (b) photon added coherent state are shown. Here, we have
choosen coherent state parameter $\alpha=0$. Therefore, (a) actually
corresponds to vacuum state, which is classical and (b) corresponds
to Fock state $|1\rangle$, which is nonclassical. }
\end{figure}

Here we aim to briefly mention the concepts that are used to identify
(characterize) a radiation field having nonclassical characteristics.
We have already mentioned that $P$-function cannot be measured directly.
The same is true for Wigner function, which is also a quasi-probability
distribution, and a quantum state in the quadrature phase space $\left(q,p\right)$
is defined as 
\begin{equation}
\begin{array}{lcl}
W\left(q,p\right) & = & \frac{1}{2\pi\hbar}\int d\xi\langle q-\frac{\xi}{2}|\rho|q+\frac{\xi}{2}\rangle\exp\left(\frac{i\xi p}{\hbar}\right).\end{array}\label{eq:wigner}
\end{equation}
Negative values of it characterizes nonclassical state. See Fig. \ref{fig:fig2wigner},
where we have plotted Wigner function of coherent state in Fig. \ref{fig:fig2wigner}
(a) and the same for PACS in Fig. \ref{fig:fig2wigner} (b). Once
can clearly see that Wigner function of coherent state is always positive
as the state is classical, but the Wigner function of PACS is negative
in some places, and the observed negativity works as witness of nonclassicality
for PACS. Thus, Wigner function can be used as a witness of nonclassicality,
but there does not exist a general procedure for the measurement of
Wigner function. More precisely, there are a few papers \cite{lutterbach1997method,banaszek1999direct,bertet2002direct}
that report the determination of nonclassical characteristics of radiation
field (negative regions in the Wigner function) by direct measurement
of the Wigner function, but the methods adopted there work for particular
cases only and there does not exist any general method for the direct
determination of Wigner function. So we characterize nonclassicality
through other operational criteria for nonclassicality. Several experiments
are routinely performed to characterize nonclassical light. A nice
list of early experiments on nonclassical light is provided in Table
1 of Ref. \cite{teich1989squeezed}, which also provide a lucid introduction
to the experimental techniques used in those pioneering experiments.
Without elaborating on all the techniques here, we may mention that
in one of the pioneering experiments on quantum optics, in 1977, Kimble,
Dagenais, and Mandel demonstrated antibunching in resonance fluorescence
\cite{kimble1977photon}. Subsequently, in 1983, Short and Mandel
\cite{short1983observation} used the resonance fluorescence again
to demonstrate the existence of sub-Poissonian photon statistics,
and in 1985, quadrature squeezing of vacuum was shown using non-degenerate
FWM process in Na atoms \cite{slusher1985observation} by using an
idea proposed in 1979 by Yuen and Shapiro \cite{yuen1979generation}.
Thus, antibunched states were prepared in 1977, but it took another
8 years to generate squeezed state.\textcolor{red}{{} }More recently,
squeezed state generation in optomechanical systems \cite{pirkkalainen2015squeezing,rashid2016experimental}
and higher order correlations in various states of the radiation field
\cite{allevi2012measuring,avenhaus2010accessing,hamar2014non} have
also been reported. The set of experiments indicates the possibility
of characterizing higher order squeezed light. Further, several closely
related experiments having applications in realizing various schemes
for quantum communication have also been performed in the recent past
(see \cite{pathak2016optical} and references therein).

Here, it would be relevant to note that realization of BB84 and various
other protocols of quantum cryptography requires single-photon sources,
but there does not exist any on-demand single-photon source. However,
there exist various approximate single-photon sources, and all of
them are expected to show antibunching \cite{pathak2010recent}. As
a consequence, it has become quite relevant to check whether a given
state of light is antibunched. This characterization is usually done
using the famous Hanbury Brown and Twiss (HBT) experiment \cite{brown1956test}.\textcolor{blue}{{}
}In this experiment, the light from a source is made to fall on a
beam-splitter, and two detectors (${\rm D_{1}}$ and ${\rm D_{2}}$)
are placed in the two output ports of the beam-splitter at equal distance
from the beam-splitter. The outputs of the detectors are connected
to a correlator or coincidence counter, which records both the number
of counts and the time delay between the clicks on two different detectors.
Specifically, the plots generated from the correlation counts reveal
the probability of simultaneous clicks of two detectors compared to
the consecutive clicks on the same detectors for different values
of delay. 

This can generate three possible scenarios captured in the correlation
function. In the first case, when the probability of simultaneous
clicks is greater than that of the consecutive clicks (clicks with
a delay), the state of light is considered as bunched. On the contrary,
when simultaneous clicks are less probable than the consecutive clicks,
the light is characterized to be in the antibunched state. The correlation
function for the first (second) case shows a peak (dip) at zero delay
time. The third case is that of the equally probable consecutive and
simultaneous clicks, which corresponds to a laser source (coherent
state). The correlation function remains unchanged for every value
of delay time. A conventional light source (namely, filament bulb)
produces bunched states of light as they usually generate multiphoton
pulses. It is worth pointing out here that the photons are simultaneous
only if they reach the detectors within the resolution time (dead
time) of the detectors, which is about 20-50 ns.

This coincidence-count-based scheme can be easily modified to design
a scheme for detecting quadrature squeezing. Interestingly, in contrast
to antibunching, squeezing is a phase sensitive property. This is
why a strong laser beam (local oscillator) is made to incident on
the second input port of the beam splitter used in HBT experiment.
When the input light incident on the first input port of the beam
splitter mixes with a strong beam (local oscillator) at the beam splitter,
at the output the difference of the current from both the detectors
is used to observe squeezing by varying the phase of the local oscillator.\textcolor{red}{{}
}When the local oscillator of the same frequency as the beam under
consideration is used, it is referred to as the homodyne detection,
while different frequency corresponds to heterodyne detection. 

There is one more interesting phenomenon associated with beam splitter,
i.e., when two single photons reach a beam splitter simultaneously
due to their bosonic nature both of them take the same output port
of the beam splitter. This can be verified by checking the photon
number detection in both the output ports. This phenomena is known
as Hong-Ou Mandel effect.\textcolor{red}{{} }

\section{Applications of nonclassical light\label{sec:Applications-of-nonclassical}}

In this section, we aim to discuss applications of different types
of nonclassical light (e.g., squeezed, antibunched and entangled states
of light) with a brief introduction to the corresponding history.
To begin with, let us briefly review the early history of squeezed
state and its modern applications. 

\subsection{Squeezed state and its applications\label{subsec:Squeezed-state-and} }

We have already mentioned that squeezed state was discovered by Kennard
in 1927 \cite{kennard1927quantenmechanik}. An extremely interesting
history of this discovery can be found at Sec. 4 of \cite{nieto1997discovery}.
Here we would like to narrate the story in brief. Earle Hesse Kennard
was an assistant professor at Cornell University, and he was a granted
a sabbatical in 1926. In October 1926, he reached Institut ${\rm f\ddot{u}r}$
Theretische Physik, University of ${\rm G\ddot{o}ttingen}$, where
Max Born used to work at that time. There Kennard learned the matrix
mechanics of Heisenberg and the wave mechanics of Schrodinger. This
was a very productive period in physics, during this time, Heisenberg
submitted his famous paper on uncertainty relations paper and went
to Copenhagen to work with Bohr. Almost immediately after that (on
March 7, 1927, Kennard also reached Copenhagen to work with Bohr.
At Copenhagen, he completed the manuscript \cite{kennard1927quantenmechanik}
that reported the discovery of the squeezed states. In that paper
he acknowledged the help received from Bohr and Heisenberg. It was
great contribution, but its importance was not properly understood
until experimental quantum optics took shape. Although squeezed state
was discovered in 1927, the term ``squeeze'' was coined much later
in 1979 in the context of increased sensitivity of an antenna designed
for the gravitational-wave detection \cite{Hollenhorst1979quantum}.
It may be interesting to note that in \cite{Hollenhorst1979quantum}
terms like squeeze operator and squeeze factor were used, but squeezed
state was not used explicitly. Further, more interestingly, in 2016,
the existence of gravitational wave has been confirmed in the famous
LIGO experiment using squeezed states and a method in the context
of which the term squeeze appeared in the world of quantum optics
\cite{aasi2013enhanced,grote2013first}. The essential physics of
using squeezed state for the detection of gravitational wave was known
for long. The activities in this direction were actually initiated
in 1980s \cite{schechter1986searching}, through the seminal proposal
of Caves \cite{caves1981quantum}. In a lucid manner, the procedure
for gravitational wave detection can be visualized as follows. Consider
that we have a Michelson interferometer and a laser as a source of
light. Now, if a gravitational wave originated due to supernova explosion
or black hole merging causes vibration of a mirror of the Michelson
interferometer, then that would cause modulation of the reflected
laser light from that mirror and consequently the interference pattern
would be changed. The change in interference pattern can be detected
by the appropriate detectors, but in the usual situation (i.e., when
no squeezed light is used), the sensitivity of the interferometer
would be limited by the fluctuations of the vacuum state entering
through the unused port of the interferometer. Specifically, sensitivity
limit arises because of two types of noise- photon counting and radiation
pressure fluctuations, which originate due to fluctuations in the
two different quadratures associated with the vacuum that enters through
the unused input port of the interferometer. To beat this sensitivity
limit, squeezed vacuum state is injected into the system through the
otherwise unused port \cite{aasi2013enhanced,grote2013first,teich1990squeezed}.
This would reduce one of the above mentioned noises, depending upon
which quadrature of the squeezed vacuum state is squeezed. Following,
similar argument, sensitivity of other devices can also be improved
using squeezed light. To be precise, quantum fluctuations limit the
sensitivity of measuring devices. However, this quantum uncertainty
can be circumvented by using the quiet component (squeezed quadrature,
say $X$ quadrature) of the squeezed state of a radiation field and
by using a detection technique that is insensitive to the noise present
in the other quadrature ($Y$ quadrature in our case) \cite{teich1990squeezed}.
Another interesting application of squeezed state is an optical waveguide
tap which was introduced by Shapiro in 1980 \cite{shapiro1980optical}.
In an optical waveguide tap, squeezed state is sent through a waveguide
which is used to tap another waveguide that carries the actual information.
The use of squeezed state helps us to obtain a very high signal to
noise ratio (SNR).

\begin{figure}[H]
\begin{centering}
\includegraphics[scale=0.6]{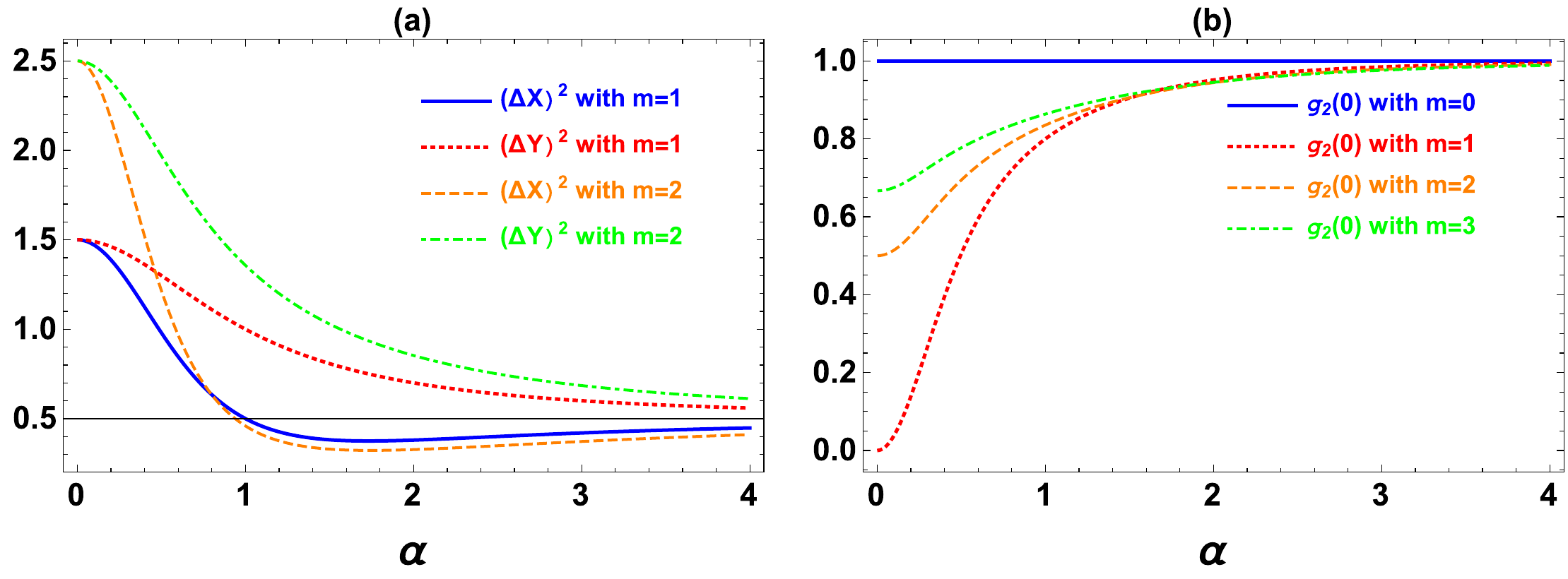}
\par\end{centering}
\caption{\label{fig:-HOM-HOA}(Color online) The variation of witnesses of
squeezing and antibunching is shown for the photon added coherent
state with the coherent state parameter $\alpha$ in (a) and (b),
respectively. Here, $m$ corresponds to the number of photons added
to the coherent state.}
\end{figure}

Squeezed state can also be used for teleportation of coherent states
\cite{furusawa1998unconditional} and for continuous variable quantum
key distribution (CVQKD) \cite{hillery2000quantum} in particular,
and quantum communication in general \cite{braunstein2005quantum,braunstein2012quantum}.
Out of these interesting applications, CVQKD needs special mention
as it can provide unconditional security to the transmitted information.
Detail description of the scheme proposed by Hillery can be found
at \cite{hillery2000quantum}, here we briefly note that in Hillery's
work, a quantum state is viewed as a point in a phase space defined
by $X$ and $Y$ quadratures (axes) and the point is surrounded by
an error box. The error box would represent the quantum fluctuation.
For coherent state, it would be a circle of radius $r$, whereas for
a minimum uncertainty squeezed state (a state with $(\Delta X)^{2}(\Delta Y)^{2}=\frac{1}{4},$
but $(\Delta X)^{2}\neq(\Delta Y)^{2}$) it would become an ellipse
and thus allow us to define one quadrature in a precise manner at
the cost of precision in\textcolor{red}{{} }the other quadrature (cf.
Fig. \ref{fig:-HOM-HOA} a, where squeezing in $X$ quadrature is
witnessed for PACS defined in Eq. (\ref{eq:pacs}) through the reduction
of $(\Delta X)^{2}$ below $\frac{1}{2}$, which is the value of $(\Delta X)^{2}$
for the coherent state). Now if Alice wants to distribute a key to
Bob in a secret manner, using squeezed state, she may follow the following
strategy suggested by Hillery.\textcolor{red}{{} }The sender (Alice)
and receiver (Bob) divide both axes into segments (bins) of equal
sizes, which are essentially less than $\frac{1}{2}$ in size. Each
bin corresponds to a symbol, and the number of allowed bins depends
on the length of the major axis. As a specific case, we may consider
that only 2 bins are allowed and they correspond to 0 and 1, which
can be chosen randomly on either of the axes. Specifically, Alice
can encode a bit value 0 in two different ways, i.e., she can prepare
a quantum state centered at $X$-axis ($Y$-axis) in the first bin
squeezed in $X$ ($Y$) quadrature. This state has well defined $X$
($Y$) value and $Y$ ($X$) value is poorly defined. Independently,
Bob is also allowed to measure one of the quadratures at random using
Homodyne detection technique discussed in Sec. \ref{sec:Characterization-of-nonclassical}.
At the end of this step, Alice and Bob reconcile the choices of quadratures
they have made to encode and measure. They discard all the cases,
except where they have made the same choice. Using this method Alice
and Bob share a symmetric key, whose security is ensured by checking
half of the shared symmetric key. 

As this review is not focused on squeezed states alone, we could not
describe all the aspects and applications of the squeezed state. Interested
readers may obtain more information about its interesting features
and applicability in classic reviews \cite{walls1983squeezed,loudon1987squeezed,nieto1997discovery}
and a few relatively new reviews \cite{dodonov2002nonclassical,andersen201630}.

\subsection{Antibunched state and its applications\label{subsec:Antibunched-state-and} }

To lucidly visualize the phenomenon of antibunching, we may note that
sometimes (in some states of radiation field) photons prefer to travel
alone (one by one) and that leads to antibunching. Specifically, if
we find a state of light in which photons prefer to travel alone in
comparison to traveling with another photon then we refer to the state
of light as antibunched and the corresponding phenomenon as the photon
antibunching (see Chapter 8 of \cite{ghatak2015light}). Antibunched
light is nonclassical light as antibunched states don't have any classical
counterpart. This nonclassical state has been investigated since long
time \cite{kimble1977photon,loudon1980non,teich1990squeezed}. Recently
on-chip generation of antibunched light has been reported in Ref.
\cite{khasminskaya2016fully}. Similar to the notion of antibunching,
a notion of bunched states of light may be introduced as a state of
light in which photons prefer to travel in the company of other photons.
Sunlight and light received from the lamps used at home are in bunched
state. In addition, there are some sources of light (like lasers)
which neither show any preference for traveling alone nor for travelling
in groups. Such a state of light is considered coherent. A simple
experiment designed by Hanbury Brown and Twiss (HBT), who were astronomers
interested in measurement of diameter of stars can be used to determine
whether the light coming from a source is antibunched, bunched or
coherent. The experiment is briefly described in Sec. \ref{sec:Characterization-of-nonclassical}.
Usually possibility of observing antibunching is checked using the
following criterion: $g_{2}(0)<1,$ where $g_{2}(0)=\frac{\langle a^{\dagger}(t)a^{\dagger}(t)a(t)a(t)\rangle}{\langle a^{\dagger}(t)a(t)\rangle\langle a^{\dagger}(t)a(t)\rangle}<1.$
A coherent state always yields $g_{2}(0)=1$, and we say that the
light is unbunched and a thermal state gives $g_{2}(0)>1,$ which
implies a bunched state of light where photons prefer to travel together.
In Fig.  \ref{fig:-HOM-HOA} b, one can easily observe that PACS defined
by Eq. (\ref{eq:pacs}) is antibunched. Antibunched states are of
interest for various reasons. Firstly, they show a unique manifestation
of nonclassicality. To illustrate this point, we may note that Fock
states, which are considered to be most nonclassical show antibunching,
but they don't show squeezing.

Antibunching is closely related to sub-Poissonian photon statistics.
Specifically, for a short counting time, the presence of antibunching
would ensure the presence of sub-Poissonian photon number distribution
and vice versa \cite{teich1990squeezed}. The sub-Poissonian photon
statistics is already defined above through the criterion $\left(\Delta N\right)^{2}<\bar{N}.$
As for coherent (Poissonian) state, we obtain $\left(\Delta N\right)^{2}=\bar{N},$
sub-Poissonian photon statistics essentially represent a state where
fluctuations in photon number is less than that in the most classical
(coherent) state. Thus, it may be referred to as the photon number
squeezed state. In context of the applications of squeezed states,
we have already discussed how the squeezing in one of the quadrature
helps us in performing accurate measurements. Following the same argument,
we may say that the photon number squeezed states may be useful in
performing precise measurements where the intensity of the incident
beam (the number of photons present in the beam) matters. A set of
such applications is discussed in \cite{teich1989squeezed,teich1990squeezed}.
Here we briefly note that sub-Poissonian light may be used to compare
the roles of photon noise (which is reduced in the case of the sub-Poissonian
light), retinal noise and neural noise in the visual response at threshold.
Specifically, in our retina, in response to light, ganglion cells
generate and transmit neural signals to higher visual centers of the
brain using the optic nerves. The statistical nature of this signal
gets affected by photon noise, retinal noise and neural noise. By
using sub-Poissonian light, we can reduce the effect of photon noise
and thus isolate the effect of other noises (for detail see \cite{teich1989squeezed,teich1990squeezed,teich1982multiplication}
and references therein). Further, the use of sub-Poissonian light
as a stimulus in visual psychophysics may help us to understand the
process of seeing at the threshold \cite{teich1990squeezed,teich1982multiplication}.
Specifically, it may help us to understand what governs the uncertainties
that appear in the human visual response near the threshold of seeing.
In optical communication systems, there are various sources of noise,
including photon noise intrinsic to the source of light. Use of sub-Poissonian
light as a source can reduce this particular type of noise and thus
the errors caused due to this noise \cite{teich1989squeezed}. In
brief the use of sub-Poissonian light helps us to improve the accuracy
of those equipment whose sensitivity is restricted by the quantum
fluctuations in the number of photons present in the radiation field.
As mentioned above, for a short counting time, antibunching and sub-Poissonian
photon statistics are equivalent and thus, these applications of sub-Poissonian
light can also be viewed as applications of antibunched light. Further,
antibunching is reported to be useful in characterizing single-photon
sources \cite{pathak2007mathematical}.

Recently, antibunching has been reported theoretically in \cite{thapliyal2014higher,thapliyal2014nonclassical,pathak2013nonclassicality}
and experimentally in \cite{kimble1977photon,zwiller2001single,zhou2016strong,gulati2014generation,stevens2014third}.
Thus, this particular type of nonclassical light seems to be easily
achievable in many physical systems and have interesting applications
in various domains of physics. 

\subsection{Quantum state engineering \label{subsec:Quantum-state-engineering} }

Until now we have seen that there are several applications of nonclassical
states in general and nonclassical light in particular. Thus, in short,
we can say that, nonclassical states are in the heart of quantum optics.
The question is- how to generate a desired nonclassical state? There
are various ways. For example, we may find a suitable Hamiltonian
$H$ and construct corresponding unitary time evolution operator $U(t)=\exp\left(-\frac{iHt}{\hbar}\right)$
that would lead to the desired nonclassical state $|\psi_{{\rm desired}}\rangle=U(t)|\psi_{{\rm initial}}\rangle$
after evolution of a given initial state $|\psi_{{\rm initial}}\rangle$
for time $t;$ it may also be constructed by performing an\textcolor{magenta}{{}
}appropriate measurement on one of the subsystems of an entangled
system, and thus compelling the other subsystem to collapse into the
desired nonclassical state \cite{Garraway1994generation,vogel1993quantum}.
However, in practice, it is not possible to construct all Hamiltonian
or entangled states. This practical restriction encouraged scientists
to look for other routes to construct desired nonclassical states,
and the same led to a subject now known as quantum state engineering
which allows us to construct the desired nonclassical/quantum state.
In one of the pioneering works in this domain, in Ref. \cite{vogel1993quantum},
a prescription was provided for the construction any desired nonclassical
state of the radiation field using a simple single mode Hamiltonian.
This interesting approach led to many new ideas of quantum state engineering.
For example, in \cite{janszky1995quantum} Janaszky et al., provided
a recipe for the construction of a set of superposition states that
coincide with the Fock states for any practical purpose. Specifically,
it was shown that for all practical purposes, a Fock state $|n\rangle$
can be viewed as a superposition of $n+1$ coherent states having
small amplitudes. Earlier works of the same group \cite{janszky1993coherent,domokos1994role}
established that some nonclassical states can be arbitrarily well
approximated as a discrete superposition of coherent states. These
early efforts of quantum state engineering led to many recent and
interesting ideas. For example, as the state engineering allows one
to create finite dimensional states of the radiation field, the process
of generation of finite dimensional state is viewed as scissors which
can truncate the usually infinite dimensional Hilbert space into an
$N$ dimensional Hilbert space. Thus, the scissors cuts a finite dimensional
Hilbert space from the infinite dimensional Hilbert space. Such a
process of truncating the Hilbert space is referred to as quantum
scissors \cite{ozdemir2001quantum,miranowicz2004dissipation,miranowicz2014phase}
(see Fig. \ref{fig:quantum-scissors} for a feeling of the task performed
by quantum scissors). We have already noted that any finite superposition
of Fock states is nonclassical (for a review on nonclassicality of
the finite dimensional states see \cite{miranowicz2003quantum,miranowicz2003quantumII}),
thus quantum scissors usually provides nonclassical states. For example,
finite dimensional coherent states are nonclassical and they may be
produced using quantum scissors implemented with beam splitters, detectors
and mirrors as shown in \cite{miranowicz2014phase}. Further, states
produced through quantum scissors may be used for teleportation of
single mode optical states \cite{babichev2003quantum} and qudit states
\cite{miranowicz2005optical}.

\subsection{Many facets of quantum communication\label{subsec:Many-facets-of} }

In the physical implementations of various schemes of QKD and MDIQKD,
in place of a single-photon source weak coherent pulse (WCP) is used.
For example, see Fig. 1 of \cite{lo2012measurement} and Fig 3 of
\cite{abruzzo2014measurement}. When a weak coherent pulse ($|\alpha\rangle$)
is transmitted through a neutral density filter it reduces the average
photon number $|\alpha|^{2}$, without altering the photon number
distribution $P(n).$ In such cases, when $|\alpha|^{2}<1$, most
of the time there will be 0 photon in the output of WCP, whenever
there will be non-zero number of photons, it will be most likely 1
photon, the probability of obtaining the output in state $|2\rangle,$$|3\rangle$,
etc., will be negligibly small. Thus, the output of WCP can be used
as an approximated single-photon source. However, as far as the $P$-function-based
definition of nonclassical light is concerned, the output of WCP is
still in coherent state and thus it's a classical light. In the next
section, we will provide more examples of applications of classical
light, before that we would like to note that even in schemes of quantum
communication where technically we require a nonclassical state of
radiation field (Fock state $|1\rangle$), we often use classical
light (WCP) in place of that. 

The simplest and most powerful use of single photon state (Fock state
$|1\rangle$, which is definitely maximally nonclassical by our discussion
so far) appeared in 1984, when Bennett and Brassard \cite{bennett1984quantum}
proposed an unconditionally secure scheme for quantum key distribution
(QKD), which is now known as BB84 protocol. In this scheme, Alice
randomly prepares a sequence of single photon states, where each state
is randomly prepared in one of the following states of polarization
$|H\rangle,\,|V\rangle,\,|\nearrow\rangle$ and $|\nwarrow\rangle$,
where $|\nearrow\rangle=\frac{|H\rangle+|V\rangle}{\sqrt{2}}$ and
$|\nwarrow\rangle=\frac{|H\rangle-|V\rangle}{\sqrt{2}}$ represent
a photon polarized at 45$^{o}$ and 135$^{o}$ with respect to the horizontal.
She transmits the sequence to Bob, who at a later time measures the
states randomly using $\left\{ |H\rangle,|V\rangle\right\} $ or $\left\{ |\nearrow\rangle,|\nwarrow\rangle\right\} $
basis, and announces the basis used to measure a particular qubit.
If the basis used by Bob for measurement and that used by Alice are
same, Bob keeps the qubits, otherwise they discard. Now, Bob randomly
selects half of the remaining qubits as verification qubits and announces
the outcomes of those measurements. Ideally (i.e., in the absence
of any Eavesdropper (Eve)), measurement outcomes of Bob would perfectly
match with the states prepared by Alice as they have used the same
basis. Any deviation from that would indicate the presence of Eve
or noise, and if a mismatch greater than a pre-computed tolerable
rate is found, they discard the protocol, otherwise they use the rest
of the qubits (after some post-processing as key for future communication).
Uncertainty principle restricts Eve from performing simultaneous accurate
measurement using $\left\{ |H\rangle,|V\rangle\right\} $ and $\left\{ |\nearrow\rangle,|\nwarrow\rangle\right\} $
bases as the corresponding measurement operators do not commute (cf.
$\left[M_{H},M_{\nearrow}\right]\neq0,$ where $M_{H}=|H\rangle\langle H|$
and $M_{\nearrow}=|\nearrow\rangle\langle\nearrow|$ are measurement
operators from $\left\{ |H\rangle,|V\rangle\right\} $ and $\left\{ |\nearrow\rangle,|\nwarrow\rangle\right\} $
basis, respectively. As Eve does not know which qubit (photon) is
prepared in which basis, any eavesdropping effort by her would imply
measurement of some of the qubits in the wrong basis (i.e., in a basis
other than the basis in which it was prepared), and that would leave
detectable traces of eavesdropping. The security of this single photon
(nonclassical state) based scheme is unconditional as it is obtained
from the fundamental laws of physics and not from the computational
difficulty of a mathematical problem. The unconditional security achieved
is a desired feature, but it is not achievable in the classical world.
This particularly interesting feature of this scheme led to a bunch
of similar Fock-state-based (single-photon-based) schemes for various
secure communication tasks. Some of them were restricted to QKD \cite{bennett1992communication}
and some of them were extended to perform secure direct quantum communication
\cite{kak2006three,lucamarini2005secure}\footnote{In \cite{kak2006three}, the author had described the scheme as a
scheme for QKD, but a careful look into the scheme easily reveals
that the scheme proposed in \cite{kak2006three} was actually a scheme
for quantum secure direct communication.}, where a message can be communicated directly without prior generation
of keys. Some foundationally important ideas have essentially been
explored using Fock state $|1\rangle.$ Specifically, the implementation
of counterfactual measurement or interaction free measurement or Elitzur-Vaidman
bomb testing \cite{elitzur1993quantum} and Guo-Shi scheme of counterfactual
QKD \cite{guo1999quantum} requires Fock state $|1\rangle$ and thus
the nonclassical light (for a lucid description of these schemes see
Chapter 8 of \cite{ghatak2015light}). Further, in various entangled-state-based
schemes for secure communication, single photons from a sequence of
single photons prepared randomly in $|H\rangle,\,|V\rangle,\,|\nearrow\rangle$
and $|\nwarrow\rangle$ are inserted randomly in the sequence of message
qubits as verification (decoy) qubits which are subsequently measured
and compared in a manner similar to what was followed in BB84 protocol
and this strategy analogous to BB84 protocol is referred to as BB84
subroutine \cite{sharma2016verification} gives unconditional security
to those schemes. For example, in the original Ping-Pong protocol
\cite{bostrom2002deterministic} and LM05 protocol \cite{lucamarini2005secure}
of quantum secure direct communication, B92 protocol \cite{bennett1992quantum}
for QKD, quantum key agreement protocol by Chong et al. \cite{chong2010quantum},
and Shi et al.'s quantum dialogue scheme \cite{shi2010quantum} unconditional
security is derived from the use of Fock state $|1\rangle.$ Further,
there exist a few commercial products, where single-photon sources
are used. Of course, there are various commercial solutions for QKD
\cite{IDQ,TOSHIBA,MITSHUBISHI}, but a quantum random number generator
needs a special mention (cf. QUANTIS sold by IdQuantique \cite{QUANTIS})
as there does not exist any true random number generator in the classical
world, although it's required for various applications including casinos.
Working of a quantum random number generator is simple. Let's send
a single photon (i.e., Fock state $|1\rangle$) through a 50:50 beam
splitter; post beam splitter the photon will be in a superposition
state $\frac{|{\rm reflected}\rangle+|{\rm transmitted}\rangle}{\sqrt{2}}$.
Now if we put one detector along the reflected path and one along
transmitted path, this will be equivalent to measuring the superposition
state using $\left\{ |{\rm reflected}\rangle,|{\rm transmitted}\rangle\right\} $
basis, and in accordance to quantum mechanics the state will collapse
randomly to one of the possibilities, in other words, detectors will
click randomly. We may consider the click of the detector along the
reflected (transmitted) path as 0 (1), and thus obtain a truly random
sequence of 0 and 1. Thus, the applications discussed so far require
a single-photon source. However, a source that can provide on-demand$|1\rangle$
states, is not available. In other words, a source of nonclassical
light that can emit single photon as and when it is required is not
available and this is why we use either WCP (a classical light source
approximated as a single-photon source) or a heralded entangled-state-based
single-photon source \cite{pittman2002single,migdall2002tailoring}.
Entangled states are nonclassical and their use is not restricted
to the design of single-photon sources. In fact, they are used to
propose many schemes for quantum communication. Some of them (e.g.,
teleportation and densecoding) have no classical analogue and entanglement
is essential for them. For the implementation of device-independednt-quantum-key-distribution
(DIQKD), we need Bell-nonlcal states, and all pure entangles states
are Bell-nonlocal and every Bell-nonlocal states are entangled (but
the converse is not true). For another set of schemes for secure quantum
communication, entanglement is found to be useful, but not essential
(say, quantum e-commerce and quantum voting). In the following subsection,
we list a few tasks where entangled states, which are always nonclassical,
are used. 

\subsubsection{Entangled state and its applications\label{subsec:Entangled-state-and} }

It is already mentioned that entangled states, which are nonclassical
states, are essential for the realization of dense-coding \cite{bennett1992communication}
and quantum teleportation\footnote{Quantum teleportation is a very interesting process that nicely illustrates
the power of quantum mechanics. In this scheme, an unknown quantum
state is transferred using prior shared entanglement and classical
communication, but the state can not be found in the channels that
connect the sender and the receiver.} of an unknown quantum state \cite{bennett1993teleporting} and that
of a known quantum state, which is referred to as remote state preparation
\cite{pati2000minimum}. Further, entanglement is essential for implementation
of various variants of teleportation and remote state preparation,
such as probabilistic teleportation \cite{li2000probabilistic}, teleportation
using non-orthogonal states \cite{sisodia2017teleportation}, quantum
information splitting \cite{hillery1999quantum}, joint remote state
preparation \cite{wang2013joint}, hierarchical joint remote state
preparation \cite{shukla2016hierarchical}, bidirectional controlled
state teleportation \cite{thapliyal2015applications,thapliyal2015general},
bidirectional controlled remote state preparation \cite{sharma2015controlled,thapliyal2015general},
bidirectional controlled joint remote state preparation \cite{sharma2015controlled,thapliyal2015general}.
It can be used to implement schemes for secure quantum communication,
like- Ekert's protocol for QKD \cite{ekert1991quantum}, Ping-pong
protocol for QSDC \cite{bostrom2002deterministic}, protocols for
two-way secure direct quantum communication known as quantum dialogue\footnote{Due to the similarity of this two-way communication task with a telephone,
this type of scheme is also referred to as quantum telephone \cite{wen2007secure}
and quantum conversation \cite{jain2009secure}.} \cite{nguyen2004quantum,an2005secure,shukla2013group}, and its variant
asymmetric quantum dialogue \cite{banerjee2017asymmetric}, quantum
key agreement \cite{shukla2014protocols,shukla2017semi} where two
parties contribute equally to construct a key and no one alone can
decide any bit of the key, quantum conference \cite{banerjee2017quantum},
quantum voting \cite{thapliyal2016protocols}, quantum e-commerce
or online shopping \cite{shukla2017semi}, quantum sealed bid auction
\cite{sharma2016quantum}, quantum private comparison \cite{thapliyal2016orthogonal,shukla2017semi},
quantum secret sharing \cite{hillery1999quantum}, etc. Thus, in brief,
this particular nonclassical state (entangled state) is extremely
important for realizations of various schemes of secure quantum communication,
and some of such schemes have direct applications in our daily life.
For example, voting plays most crucial role in a democratic country,
secure online shopping and fair sealed bid auction is also crucial
for today's economy. In fact, for any task related to secure quantum
communication, if there exists a single-qubit-based scheme, there
must exist an entanglement-state-based counterpart (see \cite{sharma2016comparative}
for detail).

This is also an integral part of device independent quantum cryptography
\cite{acin2006bell}, which uses entangled states with stronger correlations
violating Bell's nonlocality.

\section{Applications of classical light\label{sec:Applications-of-classical} }

The applications described in the last section may give a perception
that all the modern applications of light are primarily focused around
nonclassical light. Such a perception is not true. In today's world,
we frequently use technologies that are based on classical light.
To be specific, just note that the output of a laser is in a coherent
state, which is a classical state of light as per the definition of
noclassicality provided through the Glauber-Sudarshan P-representation.
The recognition of the fact that laser is a classical state of light,
immediately reveals so many applications of classical light to us.
For example, we use laser to read CD/DVD, to operate cataract, to
destroy enemy's airplane in war, to send an information through optical
fiber. The domain of applications of laser is so vast that it is not
only beyond the scope of this review, it is also beyond the scope
of a single review dedicated on applications of laser. This is why
several nice reviews are written on the applications of lasers \cite{daido2012review,montross2002laser,peyre1995laser,rusak1997fundamentals,lee2004recent,hahn2012laser,harmon2013applications,michel2010review,radziemski1994review,tognoni2002quantitative,bass1995laser,black1996laser,schwarz2008laser,castellini1998laser,bagger2005review}.
However, most of them are focused on a set of particular applications.
For example, elaborate separate reviews are available on the applications
of laser-driven ion sources \cite{daido2012review}, laser shock processing
\cite{montross2002laser,peyre1995laser}, laser-induced breakdown
spectroscopy (LIBS) \cite{rusak1997fundamentals,lee2004recent,hahn2012laser,harmon2013applications}
in general, and single-shot LIBS \cite{michel2010review} and quantitative
micro-analysis performed by LIBS \cite{tognoni2002quantitative} in
particular, laser plasmas and laser ablation \cite{radziemski1994review},
laser tissue welding (a particularly important process for surgery
and tissue engineering) \cite{bass1995laser}, particle size measurement
in different industries \cite{black1996laser}, non-surgical periodontal
therapy \cite{schwarz2008laser}, laser Doppler vibrometry (LDVi)
\cite{castellini1998laser}, laser hybrid welding \cite{bagger2005review}.

From the above, we can see that LIBS drew much attention of the scientific
community. Keeping this in mind, we note that LIBS is a technique
for performing atomic emission spectroscopy using a highly energetic
(short) laser pulse as the excitation source. This method for elemental
analysis is extremely fast and in this method, the focused laser pulse
usually creates a micro-plasma on the sample surface, which leads
to the atomization and excitation of the sample. Further, almost all
kinds of traditional spectroscopic techniques (e.g., UV-VIS spectroscopy
\cite{perkampus1992uv}, luminescence spectroscopy \cite{gaft2015modern},
FTIR spectroscopy \cite{smith1996fourier,movasaghi2008fourier}, X-ray
spectroscopy \cite{chastain1995handbook,van2001handbook}, Raman spectroscopy
\cite{smith2013modern} and its variants, like surface enhanced Raman
spectroscopy (SERS) \cite{schlucker2014surface}, tip enhanced Raman
Spectroscopy (TERS) \cite{yano2014tip} and coherent anti-Stokes Raman
spectroscopy (CARS) \cite{tolles1977review}) can be viewed as applications
of classical light. These spectroscopic techniques play a crucial
role in nanotechnology (cf. applications of Raman spectroscopy in
nanotechnology \cite{amer2010raman,ferrari2013raman,jorio2012raman,souza2003raman})
to sensor designing \cite{singla2015turn}, characterization of materials
\cite{ghannoum2016optical,butler2016using,bottka1988modulation,de2016techniques}
to finding out the proof of big bang obtained through the detection
of cosmic microwave background radiation \cite{dicke1965cosmic,radziemski1994review},
drug designing \cite{cooper2002optical,fringeli1992situ} to medical
imaging \cite{arridge1999optical,tuchin2006optical,tuchin2007tissue,boas2001accuracy,bushberg2011essential},
and thus, classical light plays a crucial role in all these domains
of science. Further, almost all the quantum optical experiments use
a laser (classical light) as an initial source of light (often referred
to as pump) and generate nonclassical light via subsequent interaction
and thus the properties of classical light, even play a crucial role
in the experimental realizations of devices that can be viewed as
applications of nonclassical light.

Another interesting application of classical light (laser) is in achieving
extremely low temperature through magneto optical trapping (MOT) \cite{katori1999magneto,mckay2011low,drewsen1994investigation},
which helps us in realizing BEC (Bose Einstein Condensation) \cite{anderson1995observation,myatt1997production},
a completely quantum phenomenon. In a conventional MOT, six laser
beams (which are usually prepared from the same source) intersect
in a glass cell (cf. Fig. 1 of \cite{anderson1995observation}). Further,
we may note that in Sec. \ref{sec:Introduction }, we mentioned about
the velocity of light in the vacuum, which is very high and fixed
in free space. However, inside a medium, it reduces by a factor of
$n$, where $n$ is the refractive index of the material through which
light is passing. Usually, we come across materials with reasonable
values of refractive index. For example, $n_{{\rm glass}}=1.5,\,n_{{\rm water}}=1.33,\,n_{{\rm amber}}=1.55$.
Thus, if light passes through any of these media, it will slow down,
but would still travel with a velocity of a couple of thousand\textcolor{magenta}{{}
}km/sec, which is still very high compared to the velocities we come
across in our daily life. The question is- Is it possible to further
slow down the light? The answer is yes. Techniques for generating
ultra-slow light have been developed in the last two decades. In 1998,
laser pulses were slowed to propagate with a velocity of 17 km/sec
in a BEC of Na \cite{hahn2012laser}. Subsequently, in 2000, light
was almost stopped, stored and retrieved \cite{liu2001observation}.
The exciting progress in this domain is still continuing (for a quick
review see \cite{dutton2004art,krauss2008we} and see \cite{lukin2000nonlinear}
for a very interesting work on nonlinear optics of ultraslow single
photons).

Laser is not the only classical light in use. Lights received from
conventional sources are all classical and applications like traffic
red lights to glow signs all are classical. Such applications of classical
light are in existence since the beginning of the civilization (for
a short review on uses of classical light in optical communication
during early civilizations see Chapter 19 of \cite{ghatak2015light}).

\section{Conclusion\label{sec:Conclusion}}

The world of light is fascinating, and the discussion above provides
a glimpse of this world with a focus on different applications of
classical and nonclassical light. It is shown that many fundamental
ideas of physics were obtained through the effort to understand experiments
involving light. Further, a nonchronological review of the ideas that
have led to modern applications of optics has been provided. Using
Glauber-Sudarshan $P$-function, we have classified light as classical
light and nonclassical light and have separately discussed the modern
applications of classical and nonclassical light. In the context of
classical light, major attention is given to laser, whereas in the
context of nonclassical light, focused attention has been given to
the applications of squeezed, antibunched and entangled states of
light. Applications of single photon states have also been discussed.
As the focus of the review is modern applications of classical and
nonclassical light, we have restricted us from the detail discussion
of some closely related phenomena which arise mostly because of properties
of optical material (in some sense which is the case with the nonlinear
optics, too). Specially, we have not discussed negative refractive
index (NRI) materials \cite{shelby2001experimental,smith2004metamaterials,ramakrishna2005physics}.
We have not also discussed various types of lasers, optical fibers
and schemes of fiber optic communication. However, a set of excellent
reviews are already available in these topics. The domain of applications
of both classical light and nonclassical light is so broad that it
is almost impossible to do justice to every aspect of it. Naturally,
this review cannot also do justice to every application of light.
Still an effort has been made to lucidly introduce the readers with
the difference between classical and nonclassical light, the ideas
that led to this distinction, and the applications of these two types
of light. We conclude the review with a hope that this review will
show the link between various ideas of optics, and motivate the readers
to go through the more focused works on the applications of their
interest.\\
~

\textbf{Acknowledgment}: AP thanks Department of Science and Technology
(DST), India for the support provided through the project number EMR/2015/000393.
He also thanks Kishore Thapliyal, S Aravinda and J Banerji for their
interest in the work and Kathakali Mandal for drawing the cartoon
used in this paper. AG thanks the National Academy of Sciences India
(NASI), for supporting the present work through the M N Saha Distinguished
Fellowship.

\bibliographystyle{unsrt}
\bibliography{biblio}

\end{document}